\documentclass[useAMS,usenatbib,usegraphicx]{mn2e}
\usepackage{color}
\usepackage{multirow}
\title[$\eta$ Carinae multispectral]{A multispectral view of the periodic events in $\eta$ Carinae\thanks{Based partially on data collected at the OPD-LNA/MCT}
\thanks{Based partially on data collected at ESO telescopes} 
\thanks{Based partially on data collected at Casleo Observatory}
\thanks{Based partially on data collected at Magellan Telescopes}
\thanks{Based partially on data collected at CTIO}
}
\author[A. Damineli, et al.]{A. Damineli$^1$\thanks{e-mail:damineli@astro.iag.usp.br}, 
D. J. Hillier$^{2}$, M. F. Corcoran$^{3,4}$, O. Stahl$^5$,   J. H. Groh$^1$, J. Arias$^8$,
\newauthor
M. Teodoro$^1$,  N. Morrell$^{9}$,  R. Gamen$^{7}$,  F. Gonzalez$^7$, N.V. Leister$^1$,  H. Levato$^7$,
 \newauthor
 R. S. Levenhagen$^1$,  M. Grosso$^7$, J. F. Albacete Colombo $^6$,  G.Wallerstein$^{10}$\\ 
$^1$Instituto de Astronomia, Geof\'{\i}sica e Ci\^encias Atmosf\'ericas, 
Universidade de S\~ao Paulo, Rua do Mat\~ao 1226,\\ Cidade Universit\'aria,
 S\~ao Paulo, 05508-900, Brazil\\
$^2$Department of Physics and Astronomy, University of Pittsburgh,
 3941 O'Hara Street, Pittsburgh, PA 15260, USA \\
$^3$CRESST and X-ray Astrophysics Laboratory, NASA/GSFC, 
Greenbelt, MD 20771, USA\\
$^4$  Universities Space Research Association, 10211 Winconsin Circle, Suite 500 Columbia, MD 21044, USA\\
$^5$ZAH, Landessternwarte, K\"{o}nigstuhl 12, D-69117 Heidelberg, Germany\\
$^6$Facultad de Ciencias Astronomicas y Geofisicas de La Plata (FCAGLP)\\
$^7$Complejo Astronomico El Leoncito, Casilla de Correo 467, San Juan, Argentina\\
$^8$Departamento de Fisica, Universidad de La Serena, Chile\\
$^{9}$Las Campanas Observatory, Carnegie Observatories, Casilla 601, La Serena, Chile\\
$^{10}$Department of Astronomy, University of Washington, Seattle, WA 98195, USA\\
}

\begin{document}
\voffset=-3.0pc 

\date{Accepted XXXX XXX XX. Received YYYY YYY YY; in original form ZZZZ ZZZ ZZ}

\pagerange{\pageref{firstpage}-\pageref{lastpage}} \pubyear{2007}

\maketitle

\label{firstpage}

\begin{abstract}
A full description of the 5.5-yr low excitation events in $\eta$ Carinae is presented.
We show that they are not as simple and brief as previously thought, 
but a combination of two components. The first, the {\it slow variation} component, is revealed
by slow changes in the ionization level of circumstellar matter across the 
whole cycle and is caused by gradual changes in the wind-wind collision shock-cone 
orientation, angular opening and gaseous content. The second, the {\it collapse} component, 
is restricted to around the minimum, and is due to 
 a temporary global collapse of the wind-wind collision shock. 
High energy photons (E $>$ 16 eV) from the companion star are strongly shielded, 
leaving the	Weigelt objects at low ionization state for $>$6 months.
High energy phenomena are sensitive only to the {\it collapse}, low energy only to the 
{\it slow variation} and intermediate energies to both components. Simple eclipses 
and  mechanisms effective only near periastron (e.g., shell ejection or accretion onto 
the secondary star) cannot account for the whole 5.5-yr cycle.

We find anti-correlated changes in the intensity and the radial velocity of P Cygni 
absorption profiles in Fe\,{\sc ii}~$\lambda6455$  and He\,{\sc i}~$\lambda7065$  lines, indicating  that
the former is associated to the primary and the latter to the secondary star.
We present a set of light curves representative of the whole spectrum, useful for monitoring the 
next event (2009 January 11). 
 
\end{abstract}

\begin{keywords}
stars: general - stars: individual: eta Carinae - stars: binary.
\end{keywords}

\section{Introduction}\label{introduction}
 $\eta$ Carinae is one of the most luminous and massive stars in the Milky Way. It underwent episodes 
of large mass ejections in recent centuries, one of them creating the Homunculus bipolar flow
 with  $\sim$12~M$_{\odot}$ \citep{SGH03_IR}. It continues to lose mass at a rate of $\sim$10$^{-3}$ $\dot{\rm{M}}$ yr$^{-1}$ through a stellar wind \citep{hillieretal01}, while intervening gas and dust precludes a clear view of the central source by ground-based observations. 

Ground-based spectra show a mix of narrow, broad, permitted and forbidden emission lines 
\citep{HA92_eta,b29}, some of them displaying P Cygni absorption profiles.. 
A comprehensive description of the spectrum is found in  \citet{daminelietal07}, hereafter Paper I, 
and in the references  cited in that paper. 
The paradoxical aspect of the spectrum is the presence of lines from high and low energy states.
On the basis of  truly periodic variations of the lines, previous work was able to show  that the central 
source is a binary star.  The primary star is colder and more luminous and the invisible companion is hotter 
and fainter \citep{b5}. A tremendous wind-wind collision (WWC) was revealed from the X-ray light curve and 
spectra \citep{b2, b70, henley08}.

High resolution images from the ground and from space showed the existence 
of condensations, named Weigelt objects \citep{weigelt86}, that are the main source of  narrow lines 
and of an extended stellar wind, from where broad lines are formed \citep{b6}. 

Although we have a basic picture of the system, many details are not yet understood.
This is because the spectrum is incredibly complex and variable. Moreover, the observational 
properties of variations in lines and continua  have not been presented previously.
Some key features, like the permitted broad emission line He\,{\sc ii}~$\lambda$4686 \citep{b20}
have just been discovered  in this star, even if it is one of the most frequently observed in the
entire sky. This line eluded detection for  more than 50 years, due to its faintness (EW$<2$~\AA) 
and transient  appearance just before minimum. In spite of being faint, it reveals a huge 
reservoir of high energy photons. This line  originates
 close to the central source, but the precise location and mechanism 
 are unknown \citep{b15,SokerBehar06}. Temporal variations of the lines
 are an important tool to understand the system, but these have not been 
 fully explored.
 
 
 Extensive studies have also been made of the variability of H$\alpha$.
 \citet{kd05} examined the variability of HST STIS observations of H$\alpha$ and
 showed that the profile did not repeat at the same phase. In particular, a flat topped 
 H$\alpha$ profile was observed during the 2003 event, but this was not seen during
 the 1998/9 event. Extensive UVES observations of the SE lobe were also undertaken
 for the 2003 event \citep{b19,Weis05}. As the light is reflected by the Homunculus, this allows the events
 to be observed from different directions. Reflected variability was seen, although the line profiles are
 different from what we see  in the direction of the central star. This confirms that the event is not spherically
 symmetric, a property that could be interpreted as either a latitude dependent
 shell event, or a consequence of the different viewing angles relative to the
 binary orbital plane.  A consistent quantitative interpretation of these
 data set remains to be done.
 
In Paper I 
we presented  the fundamental parameters of the 5.5-yr cycle. They are the period 
length P~=~2022.7$\pm1.3$ d,
 the phase 0 on  $T_0=$JD 2,452,819.8 ( defined by the disappearance of He\,{\sc i}~$\lambda6678$ 
narrow component), and the high stability of the period during the last 60 years, only understandable 
in a binary scenario. However, the events are not simple eclipses, and to decipher their nature
we need to examine the temporal behavior of the permitted and forbidden lines, 
and their associated features (broad, narrow and P Cygni components).

  The paper is organized as follows. 
In section \S \ref{data} we present the source and quality of data; 
\S \ref{specinterp} we present a general view where individual emission lines originate;
   in \S \ref{binary} we present the general framework of the binary model to interpret the data;
   in \S \ref{min} we discuss the bimodal nature of the event; 
in \S \ref{hei10830} we discuss the peculiarities of the He\,{\sc i}~$\lambda10830$ line; 
in \S \ref{multispectral} we present a multispectral view of the  {\it collapse} component; 
and finally, in \S \ref{discus}, we set forth a general discussion and our conclusions.

\section{Data and measurements}\label{data}
In Paper I we described the data acquisition and measurement, so only complimentary information 
is given here. Most of the data were taken at the Pico dos Dias Brazilian 
Observatory, with additional data taken at other South American Observatories in Argentina and Chile. 
 Spectra were normalized to the underlying stellar continuum 
and the measurements performed in the standard way with the IRAF package.  
Broad line emission profiles were separated from the narrow components 
 and their  equivalent widths  were measured by direct integration under the line profile, since 
their complexity prevented fitting by standard functions.

The lines reported here cover a wide range of excitation energy and lie in the 
optical and near-infrared windows. 
Except in the violet, the stellar continuum has signal-to-noise ratio generally S/N$>$100.
Close to [Ne\,{\sc iii}]~$\lambda3868\,$\AA\ the S/N is smaller than near the other 
spectral lines, but this line is strong enough outside the minimum to produce accurate 
equivalent widths (EqWs). We do not present individual measurements 
and their associated errors, because this would require long tables and unnecessarily pollute 
the plots. Instead, we display average error bars in the figures. 
As a general rule, errors are  $\sim$5 percent for EqW$>$1~\AA\ and $\sim$ 50 m\AA\ for 
fainter lines.  The best way to evaluate the statistical errors is by looking to the 
smoothness of the line intensity curves. 
As a matter of fact, the real uncertainties are dominated by systematic 
errors and these are difficult to assess. They are caused by the extreme richness  of the 
spectrum, which makes it difficult to define the stellar continuum, and by
line deblending procedures. Systematic errors are relatively unimportant 
in the present work, as we are looking for patterns in the temporal variability. 

The spectral resolution varies from 15 to 50 km $s^{-1}$ and, since the 
radial velocity is derived from integration over many pixels, the typical uncertainty is 
$\sim$1/10 of the spectral resolution for a single pixel (better than 5 km $s^{-1}$)
Radial velocities are in the Heliocentric reference system.

\section{General interpretation  of the spectrum}
\label{specinterp}

The ground-based spectra utilize a relatively large aperture, and thus sample
both the central star, and different emitting regions in the circumstellar
envelope (e.g., the equatorial disk, the Homunculus and the Little Homunculus). 
The different line contributions can be readily identified.
First, there are the broad wind lines (H, Si\,{\sc ii}, Fe\,{\sc ii}), arising directly from the central star,
which sample a large fraction of the wind.
The broad line spectrum is similar to the P~Cygni star HDE~316285 \citep{HA92_eta,hillieretal98}. 
Conversely,
the P~Cygni absorption components are only formed on our side of the wind, in the 
line-of-sight towards the primary star. Second, the broad components 
of He\,{\sc i} emission lines, once thought to be related directly to the primary star,
are now thought to be excited by the UV radiation of the companion \citep{HGN06_UV}
and arise in the bow-shock/wind region between the two stars \citep{nielsenetal07}.
Third, there are the narrow nebular lines which arise mainly in one 
of three Weigelt objects \citep{b6}. Forbidden lines display other components,
 in addition to the narrow ones, such as the blue displaced shoulders. They seem 
to come from an extended region around the central star. They peak at $-$250\,km\,s$^{-1}$
in close coincidence with the speed of the gas in the radio spot at $\sim$1.1 arcsec NW 
from the central knot \citep{duncanetal97,b63} and both emissions could be physically connected.
Fourth, HST observations have revealed the existence of 
a `Little Homunculus' which also has its own emission spectrum \citep{Ish03_lit_hom}. Fifth,
there is intrinsic emission from the Homunculus and equatorial disk, recognized by their 
radial velocity. Sixth,
the stellar spectrum can be seen in reflection off dust within the Homunculus.
Unfortunately, in ground-based spectra, the relative contributions of these
components to the observed spectrum may change with time, and in a manner
not directly related to the binary orbit. 

\citet{HA92_eta} showed that the nebular line emitting regions
suffer significantly less extinction than does the central star. Indeed \citet{hillieretal01} 
estimated that the visual extinction to the central star was 7 mag 
in 1998 March, while \citet{verneretal02,b62}  used 0.5 mag for the
visual extinction in their analysis of the Weigelt blob spectra observed in 1998 and 1999.
It is this difference in visual extinction that causes the nebular spectra to be so
bright, relative to the stellar spectrum, in ground-based spectra.

Recent HST observations have shown that the central star has brightened by
a factor of 3 between 1998 and 2003.7 \citep{DGH99_eta_bright, Martin04},
and has continued to brighten up to 2006.5 \citep{b78}.
Since the stellar spectrum has not undergone 
marked changes, the simplest explanation is that the circumstellar extinction has
declined.  Ground-based photometry also shows changes, although of smaller amplitude \citep{b22}, 
due to dilution by Homunculus' emission.
Because of the variable extinction and the scattered radiation in the Homunculus, 
the analysis of ground-based lines is complicated. 
During the last 25 years some lines in ground-based spectra have weakened. The
observed changes could be due to intrinsic changes in the knot properties,
changes in the primary or in the secondary star, or decreasing circumstellar extinction. The
later would cause a weakening of the Weigelt nebular lines relative to the scattered stellar continuum.
 
In the case of He\,{\sc i}~$\lambda$10830  \citep{b280}, the equivalent width is decreasing faster than 
the stellar continuum is increasing, indicating that dust destruction/dissipation is not the only
cause of the variations. This
result is robust, in the sense that the flux calibration is straightforward in the near-infrared; 90 per cent of the 
energy is in the stellar continuum and the line emission comes from a well defined central knot. For lines in the 
optical range the situation is less clear, since the contrast
between the central knot and the nebula is lower.
In any case,  during an event,  extinction variations appear to be relatively small.
 Further, line ratios of neighboring nebular
lines can be used, limiting the effects of circumstellar extinction. 

Another issue of importance is the nature of the Weigelt objects which
are believed to have been ejected sometime after 1890 - either in the
smaller eruption that took place around 1890 \citep{SMG04_UV1}
or perhaps as late as the 1930's \citep{dorlandetal04}. Given their recent origin
it would not be surprising if the Weigelt spectra were undergoing
significant changes over the last 50 to 100 years. The illuminating
flux is decreasing  as the blobs move further away from the ionizing
source (ignoring possible changes in the primary and secondary stars), 
and the physical conditions (e.g., density and size) of the blobs is probably changing.
The influence of internal dust extinction within the blobs may also be changing,
as might be the differential extinction between the blobs and the central source.

The first known and convincing evidence for high excitation lines
in the spectrum of $\eta$ Carinae comes from spectra taken from 1944 through
1951 by \citet{b9}. Spectra taken in March 1938 do not clearly show He\,{\sc i}
\citep{HK05_old_spec}, and as discussed by \citet{b8}, spectra taken
before 1920 do not show the high excitation lines. This could be a
confirmation of the later ejection date for the Weigelt objects, or
simply reflect that their properties were very different from those today.
It is even possible/probable that the gas giving rise
to the narrow Fe\,{\sc ii} and [Fe\,{\sc ii}]  lines in the early 1900's
is not the same gas giving rise to the observed emission lines today. It should
be recalled that the `Little Homunculus' was ejected around 1890
\citep{Ish03_lit_hom},
and presumably this ejected material had a substantial influence on
the integrated observed spectrum for some time after it was ejected.
 In 1893 $\eta$~Carinae showed an F-type absorption spectrum, with
a significant blue shift ($-$200\,km\,s$^{-1}$) of the absorption lines, probably
indicating a shell ejection \citep{Whitney52}. That author  also notes that there were 
marked spectral changes
until 1903, while the spectrum remained relatively unchanged between 1903 and 1930.

\section{The Basic binary scenario}\label{binary}

The discovery of the strict periodicity \citep{b3} and binarity \citep{b4} 
did bring some order to explore the complex variations in lines and continuum.
While the period length is unique, the behavior of the light curve (time and duration of 
the minimum, shape of the descending and recovering branch) differs
from one feature to another. Although the ultimate cause of the event 
is the eccentric orbit, every feature is governed by a specific mechanism 
with its own characteristic timescale.  In order to discuss the different possibilities,
we use a binary model like the one presented by \citet{henley08}, 
but only as a general framework, which is 
in good general agreement with the observations, although not
completely accounting for all of them.

\begin{figure*}
\vbox {\vfil
\resizebox{17cm}{!}{\includegraphics{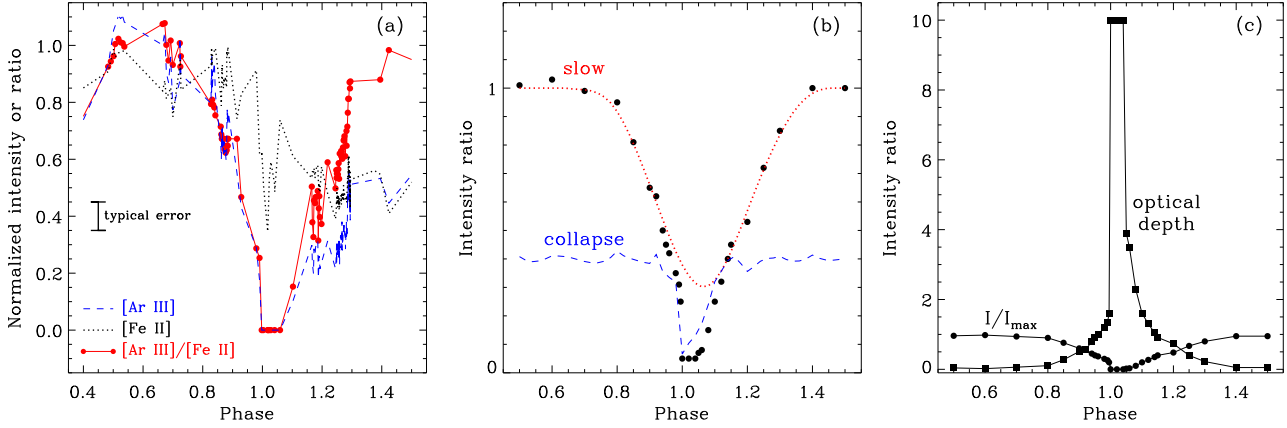}}
\caption{\label{tau} The {\it slow variation} and the {\it collapse} components seen in narrow lines.
(a)  Normalized eq. width of [Ar\,{\sc iii}]~$\lambda7135$ ({\it dashed line});
  [Fe\,{\sc ii}]~$\lambda7155$ ({\it dotted line}); and the ratio between these two narrow 
line components ({\it filled circles}). A typical error bar is shown in the middle left of the figure.
(b) Oserved intensities ({\it circles});  Gaussian fit to the "slow" component ({\it dotted line}); 
{\it collapse} component ({\it dashed line}) obtained by subtracting the Gaussian fit.
 (c)  Eq. width ratios of [Ar\,{\sc iii}]/Fe\,{\sc ii}]~$\lambda7155$  
and [S\,{\sc iii}]/Fe\,{\sc ii}~$\lambda6318$ normalized on phase  $\phi=0.4$ and averaged
 (I/I$_{\rm max}$) ({\it circles}); optical depth ({\it squares}, see text). 
} 
\vfil}
\end{figure*}

\begin{table}
\centering
\caption{\label{taut} Parameterization of the event for  [Ar\,{\sc iii}] and [S\,{\sc iii}].}
\begin{tabular}{ccc}
\hline
\multirow{2}{*}{I/I$_{\rm max}$} & Length*  & optical depth \\
 & (days) & $\tau$\\
\hline
0.90 &1214 & 0.046 \\
0.50 & 587 & 0.30 \\
0.10 & 182 & 1.0 \\
0.0  & 120 & $>>$1 \\
\hline
\hline
\multicolumn{3}{l}{* time to return to the same line intensity}
\end{tabular}
\end{table}

The system is composed of two massive and evolved stars in a highly eccentric orbit, 
the secondary companion being the main source of (hard) ionizing photons
 \citep{b4}. The secondary star is hotter and less luminous than the primary, and its wind is
 faster and less dense. The wind-wind collision (WWC) generates X-rays in 
the walls of a shock-cone bent toward the secondary star \citep{b70}. Since it 
is more transparent than the wind of the primary star, most of the 
X-rays and ionizing photons escape through it. 
Some of the photons from the secondary star penetrate the wind of the
 primary star beyond the limits of the shock-cone, producing an ionized
  cavity to the side of the companion star. The shock-cone points almost 
radially away from the primary star around apastron, but it gets twisted as it approaches 
periastron, when the orbital speed becomes comparable to that of the primary 
star wind \citep{b16}. The sudden drop in the X-ray flux has been attributed 
to the huge increase in opacity when the shock-cone opening 
leaves our line-of-sight and we see the WWC through the dense wind of the 
primary star \citep{b2}. If this is the case, the  wind of the primary star should be porous,
since during the minimum we still see a hard X-ray component with the same temperature 
as in the high excitation state \citep{b11}. We will show in this work that, even if
the orientation of the shock-cone opening is important, it is not the only or even
the most important factor to control the variability.

The observer is placed to the side of the apastron, although not necessarily aligned to the main axis of 
the orbit.  This is consistent with several observations: 
a) the He\,{\sc i} emissions are primarily blue-shifted along most of the 5.5-yr cycle;
b) the P Cygni absorptions, which must be formed on our side of the primary star, are weak 
for most of the cycle, indicating that this side is more ionized than 
the back side \citep{hillieretal01,HGN06_UV,nielsenetal07}; 
and c) if periastron were on our side of the system we should see a enhancement 
of $N_{\rm H}$ when the secondary star reaches the opposition, which would be coincident with 
the middle of the high excitation state, but which is not observed.
Further, the Weigelt objects are on our side of the system and the observation that maximum excitation 
occurs around $\phi=0.5$, when the column density ($N_{\rm H}$) to X-rays 
is low for us, indicates that the shock-cone is opened toward us 
during the high excitation state. These facts exclude models in which 
the periastron is to our side of the system \citep{b1, Kashi07}. Because of this 
fundamental disagreement, we will not comment on these works, even if they 
are in agreement with some particular aspect of the observations. In our adopted 
model, the secondary star is moving away from 
us prior to periastron passage \citep{b4,b16}. 
We assume, as usual, that the orbital plane is more or less 
perpendicular to the polar axis of the Homunculus, in order that we 
see the binary axis from an intermediate angle (neither parallel nor perpendicular), 
although there is no observational constraint to the orbital inclination.

\section{The composite morphology of the event}\label{min}

The variability of high excitation lines is reasonably well known
around the minimum. They show  a collapse in a timescale shorter than a few weeks,
 followed by a minimum a few months long. The same situation is seen in X-rays and 
in several broad-bands in the optical and near-infrared. The event, however, covers 
different timescales as we look at different spectral features.

\subsection{The {\it slow variation} component}\label{slow}

\citet{b3} and \citet{b280} showed that  He\,{\sc i}~$\lambda10830$ varies 
continuously along the 5.5-yr cycle, between minimum and maximum, with no sharp transition. 
An even more extreme example, showing almost sinusoidal variability, was reported by \citet{b7}  
for the radio cm light curve. Although this is relevant to understand the mechanism of the event, since it is 
working all along the orbit and not confined to the periastron passage,  it has not received enough attention.

In Fig. \ref{tau}a we 
illustrate variations in the EqW of [Ar\,{\sc iii}]~$\lambda7135$ (narrow component) with orbital phase for
the event \#10 (phase 0 on 1997.95). Unfortunately, we do not have photometry with the same spatial
and time resolution to derive the line flux from equivalent widths.
The asymmetry between the fading and recovering branches of the minimum is due, in part, to changes
 in the level of the continuum, but other effects are involved, since the light curve at radio wavelength is also
asymmetric \citep{b7}.  As apparent from the figure,  [Ar\,{\sc iii}]~$\lambda7135$
changes through the entire cycle. In order to compensate for changes in the continuum level we 
measured a neighboring line that suffers little variation along the event. 
We refer to the narrow line components only.
The companion line for [Ar\,{\sc iii}]~$\lambda7135$ is [Fe\,{\sc ii}]~$\lambda7155$, 
also displayed in Fig. \ref{tau}a. [Fe\,{\sc ii}] shows a  small decrease at 
$\phi=0$, but otherwise has a smooth behavior. We normalized all line intensities and ratios to unity at 
$\phi=0.4$ in order to visually compare intensity curves.  
The line ratio displays a much higher degree of asymmetry than before division 
by the neighboring line, indicating that both lines were affected by substantial variability of the stellar 
continuum along cycle \#10  (which started in 1997.95 and finished in 2003.49).

If the Weigelt objects were fully ionized outside the minimum, the line intensity curve would 
be flat in the corresponding phases, unlike what we observe.  [Ar\,{\sc iii}] shows a 
broad maximum in the range  ($\phi=0.4-0.7$), whether we look to the direct EqW measurements 
or to its ratio with the neighboring [Fe\,{\sc ii}]  line. 
These lines are strong when excitation is high (Eqw$\sim$3\AA~ for [Ar\,{\sc iii}] and EqW$\sim$10\AA~
 for [Fe\,{\sc ii}]). It seems plausible that only a fraction of Ar and Ne in the 
Weigelt objects exposed to the ionizing 
source is ionized to the second stage. We need better data, defining a clear maximum in the line intensity curve,
 to make a strong point on the partial ionization. However, it is clear that we cannot no longer say
that the events in $eta$~Carinae are brief episodes.
This seems to be in conflict with the fact that the event as seen in high energy phenomena, 
like X-rays and He\,{\sc ii}, is confined to a narrow range of phases. This apparent 
contradiction is due to the existence of two regimes in the events, as shown below.

The radio light curve at 3-cm \citep{b7} shows many similarities with the [Ar\,{\sc iii}] line intensity curve 
for event \#10; continuous variability along the whole cycle, the center of the minimum at $\sim$4.5 
months later than $\phi=0$ and the asymmetry of the minimum, and with the second branch recovering 
slower than the fading one. An important difference is that, around the minimum, the radio light curve is 
sinusoidal, but [Ar\,{\sc iii}] display a sudden drop followed by a flat minimum. 

The other three doubly ionized 
lines: [Ne\,{\sc iii}]~$\lambda3868$, [Fe\,{\sc iii}]~$\lambda4701$  and [S\,{\sc iii}]~$\lambda6312$
behave in similar way to  [Ar\,{\sc iii}]~$\lambda7135$. Since  [S\,{\sc iii}]~$\lambda6312$  was 
as densely sampled in time as [Ar\,{\sc iii}]~$\lambda7135$, we combined the two to enhance the 
signal-to-noise. In the case of  [S\,{\sc iii}]~$\lambda6312$, we divided its intensity by that of the 
neighboring Fe\,{\sc ii}~$\lambda6318$ narrow line component.
We then averaged the results for [Ar\,{\sc iii}] and [S\,{\sc iii}], after having normalized the curves to the 
unity at phase $\phi=0.4$. The combined line intensity curve is displayed as circles in Fig. \ref{tau}b. 

In order to bring more clarity to our proposed double behaviour of the event,  we  modeled it
by a Gaussian fit, excluding points deviating more than 3$\sigma$. We call the 
fit  the {\it slow variation} component. It is centered on phase $\phi=0.069$, 140 days after 
$\phi=0$ remarkably close to the radio 3-cm minimum ($\sim$130 days) and  has a full width 
half maximum fwhm = 649 days. The Gaussian fit was subtracted from the data and the difference 
is represented by a dashed curve at the bottom of  Fig. \ref{tau}b.  We call it  the
 {\it collapse} component, because of the fast drop just before $\phi=0$ and its relation with 
the minimum at high energies (X-rays and He\,{\sc ii}) as discussed in the next subsection.

What is the cause of  the slow variation changes in the doubly ionized lines? 
The intensity of the doubly ionized narrow lines depends on the ionizing flux the Weigelt 
objects receives from the secondary star. Since the binary orbit is at least 10 times 
smaller than its distance to the Weigelt objects, the change in the distance to the secondary star 
plays only a minor role in the line intensities. To get insights into the causes of the variations of the Weigelt lines, we will first interpret the radio-cm light curve, which only has a pure slow component.


Radio continuum maps \citep{b7} at 3-cm wavelengths show that the size of the 
radio source is maximum  at $\phi=0.5$, filling the Little Homunculus \citep{b63},
and then shrinking to a point source as the minimum approaches. 
If the minimum were produced by an eclipse, in which  
the ionizing source is hidden from us when passing behind the primary star,
or behind its stellar  wind, or when the opening of the shock-cone around 
the secondary star leaves our line-of-sight, there would always be 
gas being ionized in other directions, except ours. The net effect would be a 
change in the small scale structure of the 
radio map due to irregular density distribution, but not in its size and 
flux density. The facts that, during the minimum, the size of the radio map 
is reduced to almost a point source and that the flux density decreases, 
indicate that the volume of ionized gas in the circumstellar medium 
also decreases. This indicates that a fraction of ionizing photons coming the secondary 
star are impeded from escaping to the circumstellar environment, which implies that 
the size of the shock-cone opening is decreasing toward periastron and/or 
the gas density inside its cavity is increasing. The shock-cone is not
rotating as a rigid body around the center of mass, especially in regions far from its apex. 
Far from this region, the gas flow may be disturbed 
and left behind, in a spiral-like pattern, increasing the opacity for 
escaping photons.  The effect may be small for X-rays but high for UV radiation.
In this scenario,  the radio light curve is not 
expected to be symmetric with respect to the periastron passage, as is observed.

We can interpret the {\it slow variation} component of the double excited nebula lines
in the same manner as the radio light-curve. 
In this case, variations in the line intensity of doubly ionized lines represent 
changes in the optical depth toward the secondary star, as seen from the 
Weigelt objects.
In order to explore the idea that the ionization in the Weigelt objects is controlled by opacity,
let us define an optical depth, by assuming
that it is zero at maximum line intensity and increases in proportion to the decrease in the line
intensity. For example, at 90 per cent of the maximum intensity, 
the optical depth is $\tau$ = 0.046, and so on, as displayed in Table \ref{taut}.
 Fig. \ref{tau}c illustrates the change in optical depth (squares) along the
curve intensity of the doubly ionized lines (I/Imax).
We see that the optical depth has a smooth behavior 
along the 5.5-yr cycle, as expected from the {\it slow variation} component, 
suffering a sudden increase when the optical depth reaches $\sim$1.  The 
short lived {\it collapse} component corresponds to points with  $\tau > 1.$

Another way to examine the evidence for two components in the 5.5-yr cycle is by  looking for 
how long the line intensity curve stays above some intensity or above some optical depth.
As seen in Table \ref{taut}, the optical depth remains at $\tau>1$ for $\sim$6 months close to
the minimum and at $\tau<1$  for the remaining 5 years.

\begin{figure*}
\vbox {\vfil
\resizebox{17cm}{!}{\includegraphics{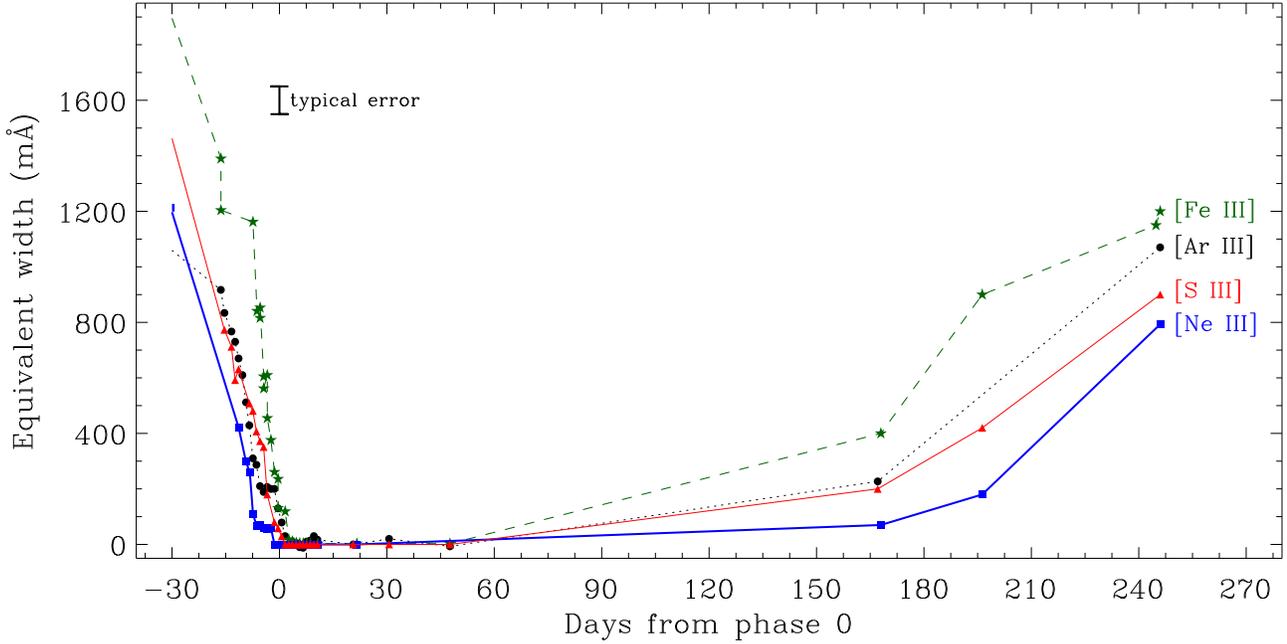}}
\caption{\label{event}The central part of the minimum for the narrow components of the high excitation lines. 
 [Ne\,{\sc iii}] ({\it squares}),  [Fe\,{\sc iii}] ({\it stars}), Ar\,{\sc iii}] ({\it circles}),  [Si\,{\sc iii}]  ({\it triangles}).
[Ne\,{\sc iii}] has the highest transition energy, fading earlier and recovering later. [Fe\,{\sc iii}]has 
the lowest energy, fading later and recovering earlier.  Ar\,{\sc iii}] and  [Si\,{\sc iii}] 
have intermediate energy. The doubly ionized lines take at least 9 months to recover the intensity they had a month before phase 0. Typical error bar is shown at the upper left of the figure.  
}\vfil}
\end{figure*}

\subsection{The {\it collapse} component}
\label{collapse}

\begin{table}
\centering
\caption{\label{delay}Time delay to $\phi=0$ for several features}
\begin{tabular}{cccc}
\hline
Feature & Time delay &  IP  & $\rho_{\rm{crit}}$ \\

pass-band & (days)&  (eV) & (cm$^{-3}$) \\
\hline
[Ne\,{\sc iii}]~$\lambda3868$			& -5&   41.0 & $7.6\times10^{6}$  \\

[Ar\,{\sc iii}]~$\lambda7135$			& +2 &   27.6  &  $5\times10^{6}$  \\
 
He\,{\sc i}~$\lambda6678$nar	        & 0&  24.6  &-\\

[S\,{\sc iii}]~$\lambda6312$			& +1.5 & 23.3  & $15\times10^{6}$ \\

[Fe\,{\sc iii}]~$\lambda4701$			& +3 & 16.2  &  - \\

[N\,{\sc ii}]~$\lambda5754$			& +3.5 &  14.5  &  $8.6\times10^{3}$   \\

X-rays					& -1&   - &- \\

He\,{\sc ii}				& +4 &54.4  &-\\

V-band					&  +20 &  -  &-\\

J-band					&  +21 &  -  &-\\

H-band					&  +15 &  -  &-\\

K-band					&  +19 &  - &-\\

L-band					&  +12 & -  &-\\

Radio 7mm			&  +29 & -&- \\

Pa$\gamma$			& (+1)  & 13.6& -\\

\hline

\end{tabular}

\end{table}

The {\it collapse} component derived by subtraction of the Gaussian fit is centered 
at $\phi=0.0299$ (60 days after  $\phi=0$), has fwhm = 189 days and covers 30 per cent 
of the area under the curve.  It starts when the doubly ionized forbidden
lines fall below $\sim$40 per cent of their maximum intensity, 
and lasts for $\sim$15 per cent of the 5.5-yr cycle.
We reinforce that these values are specific to  [Ar\,{\sc iii}] and [S\,{\sc iii}] 
and other features may result in different values, since they are affected by different proportions 
of the  {\it collapse} and the {\it slow variation} components. Moreover, these values were 
measured in the subtracted curve, which may be contaminated by defects in the 
deblending procedure.

What causes the crash in intensities in the {\it collapse} component?
The optical depth rises steeply as  $\phi=0$ approaches.
 It is unlikely due to a simple eclipse, since the wind of the primary star 
is optically thick up to many stellar radii \citep{hillieretal01} and has a gradual 
radial density profile.  At this epoch, the Weigelt objects are suddenly screened 
from high energy (E $>$ 16 eV) ionizing radiation. A possible explanation would be 
that, as the shock-cone opening changes direction around the orbit, 
the line of sight from the secondary star to the Weigelt objects crosses its walls,
entering into the high opacity region 
dominated by the primary's wind. Let us call it  the `cone eclipse'.
Interestingly all features that have the {\it collapse} component
(e.g., X-rays,  He\,{\sc ii}~$\lambda4686$, etc.) show such a fast drop.
However this does not necessarily mean that the same mechanism needs to be
invoked to explain all features --- instead we note that the time-scale for
the collapse phase is ultimately driven by the short time the
secondary spends near periastron.

 The collapse in X-rays has been attributed to the `cone eclipse'  just mentioned, 
when our line-of-sight leaves the cone opening \citep{b16}. Since the Weigelt objects 
do not lay on our  line-of-sight and both features (X-rays and doubly ionized lines) 
fade simultaneously, the same explanation cannot hold for the Weigelt lines. 
The `cone eclipse' could be attributed to the collapse of He\,{\sc ii}~$\lambda4686$ 
seen directly in the central object. However, its image reflected by the dust 
near the Homunculus south pole,  $\sim$45 degrees from  our line-of-sight,
collpases at the same time - after correcting for the extra travel time to the 
Homunculus' south pole,  \citep{b19}. In this case too the `cone eclipse'  seems 
implausible. 

Since there are many features, formed in different places of the system, which 
show a synchronized  fading, we  suggest a simple hypothesis: 
the WWC shock suffers a temporary global collapse when the stars 
are close enough to periastron.
Previous work suggested that the events in eta Car (principally changes at
the central minimum) are due to a global change in the
 WWC.  \citet{soker05}, in order to explain the fast drop 
in the X-ray curve and the long timescale of the minimum,  advocates that 
near periastron, the companion accretes from the primary's wind. 
\citet{b15} and \citep{kd02} discussed difficulties for eclipse models to account 
 for these two features in the X-ray light curve, in the He\,{\sc ii} line intensity and 
radial velocity curves. These authors suggest that the radiative and tidal forces 
of the secondary companion star induce a major disturbance 
in the inner wind of the primary, resulting in a shell ejection. Alternatively, they suggest  
that  WWC shock may become unstable because of the  large 
density  near periastron and suffer a general crash. Although our data are not 
useful to select any special mechanism,  or a particular mechanism,  
we recognize that there is a sudden  and simultaneous fading in many broad band light curves and spectral line intensities  all over the spectrum. We call it the 'collapse' component.
 
\subsection{A detailed view of the {\it collapse} component for doubly ionized lines}\label{time_scales}

In this subsection, we explore the duration of the minimum for the doubly ionized forbidden
lines and show that they disappear and re-appear in a defined sequence. Since we 
measured the EqW of the narrow component separately, they represent the
 way the Weigelt objects experience the event.

During the monitoring campaign of the 2003.5 event, we noticed that when the narrow 
component of He\,{\sc i}~$\lambda6678$ disappeared,  [Fe\,{\sc iii}]~$\lambda4701$
was still detectable and that [Ne\,{\sc iii}]~$\lambda3868$ had disappeared a few days earlier.
In Fig. \ref{event} we display the behavior of  [S\,{\sc iii}], [Ne{\sc iii}],
 [Ar{\sc iii}], and [Fe{\sc iii}] narrow components, measured in milli-Angstrom (m\AA). 
[Ne{\sc iii}], the highest excitation energy line among these four, is the first to disappear and the 
last to re-appear. [Fe{\sc iii}], with the lowest energy transition, fades later and recovers earlier. 
This could be expected naturally, without any complicated physics, if the line which fades earlier
was intrinsically much fainter outside the minimum.
But, this is not the case --- [Ne{\sc iii}] is much stronger, having reached EqW$\sim$5500 m\AA~in early 2001 
(close to the middle of  the high excitation phase) as compared to  [Fe{\sc iii}], which 
maximum was EqW$\sim$2800 m\AA. The intermediate excitation energy lines [S\,{\sc iii}] 
and [Ar\,{\sc iii}] fall inbetween the two extremes, indicating that the order 
of fading and recovering is strongly modulated by the ionization potential.
 
Fig. \ref{event} indicates that also  [S\,{\sc iii}]  and [Ar\,{\sc iii}] follow the
order {\it the line which fades earlier recovers later}. This is reinforced by the fact that
  [S\,{\sc iii}]  reaches the zero intensity  0.5 days before  [Ar\,{\sc iii}], as seen in
column 2 in Table \ref{delay}. But, this is the opposite to the trend {\it the higher the energy of the
transition, the earlier the line fades}, although we note that the ionization potentials
of the two species are very similar.  We cannot be sure that the two lines behave anomalously,
or if this is just due to the quality of our data. On the one hand, the determination of the time to 
reach the zero intensity has an uncertainty of $\sim$1 day. On the other hand, the difference 
between the EqWs of the two lines also is of the order of the errors. In addition, [S\,{\sc iii}] 
is seated on top of a variable broad Fe\,{\sc ii} line and  the nearby  stellar continuum is difficult 
to assess, and thus systematic effects may contaminate the measurements of this 
particular line.     

Since the doubly ionized forbidden lines have different ionization potentials and critical 
densities, we examined 
their correlation with the time to attain zero intensity (relative to $\phi=0$). 
The time of zero intensity was derived in the same way as in paper I 
for  the narrow component of He\,{\sc i}~$\lambda6678$: a linear fit through the descending 
branch of the line intensity curve extrapolated to zero.
In the second column of Table \ref{delay} we display the time delays for several 
spectral features, with have a typical uncertainty of  $\sim$1 day. We are focused 
on the first 6 lines in Table 1, formed in the Weigelt objects. 
Since the forbidden lines are collisionaly excited, we list the IP to get 
the ion yielding the observed transition (i.e., IP(Ne$^+$) for Ne\,{\sc iii}), but as lines 
from He\,{\sc i} are generally formed by recombination we give its IP.
We see in column 2  of Table \ref{delay}
that the time delay for disappearance has a general trend with the ionization potential 
in the sense the higher the energy earlier the line fades as seen in Fig \ref{event}. 
 As shown in that table (column  4), the time delay is not correlated  with the  critical density for line formation. 

 \begin{figure*}
\vbox {\vfil
\begin{tabular}{cc}
\resizebox{17cm}{!}{\includegraphics{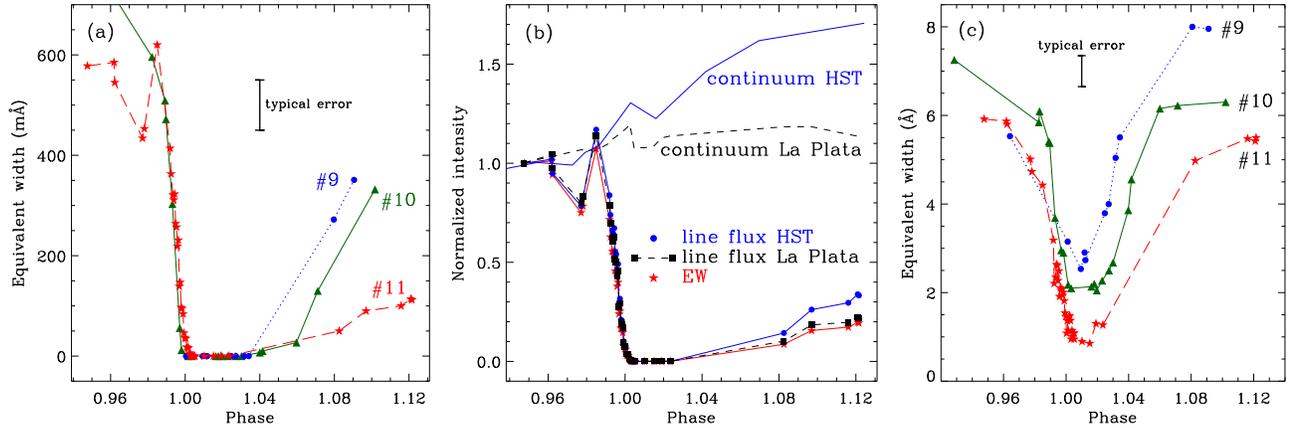}}
\end{tabular}
\caption{\label{he66} He\,{\sc i}~$\lambda6678$ line: (a) EqW of the narrow  component in the last three 
events \#9 ({\it circles}), \#10 ({\it triangles}) and \#11 ({\it stars}). 
(b)  Normalized quantities for event \#11:  continuum of the central star from 
STIS/HST HRC F550M filter ({\it upper solid curve});  $V$-band continuum of the Homunculus plus central star from La Plata 
photometry ({\it dashed curve});  narrow component line flux normalized with the (HST) stellar continuum  ({\it  circles});  
narrow line components normalized with the Homunculus continuum ({\it squares}); 
normalized EqW of the narrow line component ({\it stars}). (c) Same as in (a), but for the broad component. Typical error bars for the line measurements are shown in the upper middle of 
panels a and c. Photometric errors are smaller than the widths of the lines.
}
\vfil}
\end{figure*}

The narrow nebular lines are known to originate around the dense
Weigelt objects. If the radiation ionizing these
blobs is suddenly extinguished, the line intensities will
decay on a recombination time-scale. For He\,{\sc i}, with
$\alpha_B=2.73 \times  10^{-13}$ at 10000\,K \citep{Osterbrock89}, 
the recombination time scale is
\[
  t_{rec}=42.4 \left( { 10^6\,\rmn{cm}^{-3}\over N_{\rm e}}\right) \,\,\, \rmn{days} 
\]
where $N_{\rm e}$ is the electron density. Since He\,{\sc i} decays on a time
scale of order 5 to 10 days, the density in Weigelt objects must,
conservatively, exceed 10$^6$cm$^{-3}$.
Higher densities are possible, since it is likely that the
line variations are also governed by the timescale associated with
the reduction in ionizing flux, which in turn is associated with the
orbital time scale around $\phi=0$.
The lower limit to the electron density is consistent with that given by \citet{b62} 
who found N$_{\rm e}\sim 10^7$~cm$^{-3}$ for the high ionization region
from an analysis of Weigelt line ratios. It is consistent also with the results 
reported by \citet{b12} for the H\,{\sc ii} zones  of the Weigelt objects.

The variation in other line intensities could also be used to place constraints on the density
in the Weigelt objects. Hydrogen has a similar recombination timescale to He\,{\sc i}, 
while elements, such as N have a timescale a factor of a few shorter. On the other hand, 
elements such as O (whose ionization potential are nearly the same as H), 
are coupled to the H by strong charge exchange reactions, 
and thus their recombination timescale is determined by H. 

For a simple spherical ionization bounded nebula, the 
ionization timescale is similar to the recombination timescale \citep{Spitz78}.
As the observed recovery in line strengths takes considerably longer than the decline,
the variation in the obscuration of the ionizing source by intervening gas must be the primary factor 
determining the recovery time.

\subsection{The {\it collapse} component  in  He\,{\sc i} narrow emission lines}\label{narrow_hei}

 In Fig. \ref{he66}a we present the EqWs  of the He\,{\sc i}~$\lambda6678$ narrow line component
along the last 3 events. The label of the events are the same as in Paper I, in order that 
 \#9 has phase 0 on 1992.42, \#10 on 1997.95 and \#11 on 2003.49. There are two remarkable
features in the {\it collapse} component of this line: a) it is asymmetric and b) the post minimum 
branch has been recovering at slower pace as time goes by. 
Both behaviors could be due to changes intrinsic to the line emission region,  
or just to a temporal increase in the level of the 
stellar continuum. If we demonstrate that changes in the stellar continuum across the 
{\it collapse} component are not the cause of the asymmetry, then the secular weakening 
of the recovering branch is also not due to brightening of the central source. The 
{\it collapse} component is 
a local feature, insensitive to temporal variations in the stellar continuum flux, which occur 
on longer timescales.

In order to correct for possible changes in the stellar continuum across the minimum,
we used published photometry to derive line fluxes for the  He\,{\sc i}~$\lambda6678$ narrow 
line component along the event \#11.
Here, we are not interested in the absolute fluxes, only in the relative variations across
the core of the event ({\it collapse} component). In this way, we normalized all quantities
 to the value they had on JD~2,452,714 (represented by stars in Fig. \ref{he66}b). 
The relative line flux is obtained by multiplying the relative flux nearby continuum. by 
equivalent width.

Variations in the continuum flux are easily derived from existing photometry, since we
need only relative changes, avoiding complications involved in magnitude standardization.
Color variations along the event are negligible in the optical window (at wavelengths longer than the Balmer jump),
as indicated by the B-V color index \citep{vangenderen03, b14}. The problem is that the amplitude of 
the variations depends on the aperture used to extract the magnitudes, and the ones of published photometry 
do not match the slit aperture used to record the spectra. The slit apertures encompass a few arcsecond (1.5-4 arcsecond), as compared to $\sim$22 arcsecond used in ground-based photometry. 
Although space-based photometry is available, the extraction apertures (0.3 arcsec) are much smaller than the slit widths. Since the slit width is intermediate to  these 
two data sets, we can use them to constrain the flux variability.

The adoption of ground-based photometry would smear out the variations of 
the central star, since the extraction aperture encompasses the whole Homunculus nebula,
which is bright in the optical region.
 The line flux variations obtained in this way are a lower  limit to the one sampled 
by the slit width. In Fig. \ref{he66}b (dashed line) we show flux variations derived from photometry taken 
at La Plata \citep{b14}. The magnitude is V$\sim$5 and the errors are $\sim$0.005 magnitudes, 
in order that they are a little larger than the width of the dashed line. 
We used the $V$-band measurements, since the $R$-band is 
contaminated by H$\alpha$. The mismatch in wavelength between the line and the continuum
 is not important, since we are
dealing with relative fluxes and the color variations involving these two filters are 
very small. Relative fluxes in the $V$-band continuum, used to calibrate the line flux
 are presented by a dashed   line at the upper part of the figure, labeled as `continuum La Plata'. 

We can obtain an upper limit to the flux variations by using 
photometry from the central star, free from nebular contamination, as 
reported by \citet{b78} in their Table 5. Those authors report synthetic photometry 
derived from HST STIS imaging through the medium-band HRC F550M filter.
The magnitude for the event \#11 was $\sim$6.5 with errors smaller than 0.01 magnitudes.
Relative fluxes for the central star are presented as a  solid line in the 
upper part of Fig. \ref{he66}b, labeled as `continuum HST' and the corresponding line fluxes are 
displayed as  dots at the bottom of that figure. 
As in the case of ground-based photometry, 
errors are similar to the thickness of the line that represents the 
continuum flux.
The real relative fluxes of the He\,{\sc i}~$\lambda6678$ narrow line component are intermediate 
between the ones derived from ground and space-based photometry. 
Taking into account the errors, the asymmetry between the descending and  appears to real, and
not an artifact of variations in the continuum. On the other had, allowing for
the continuum variation does weaken
the trend that the recovery phase is changing from cycle to cycle. Frequent observations during 
the recover phase of cycle 12 may help to clarify the cycle-to-cycle variatons.

\subsection{The {\it collapse} component in the  broad He\,{\sc i}emission lines }\label{broad_hei}

In  Fig. \ref{he66}c  we display the EqW of the He\,{\sc i}~$\lambda6678$ broad line component 
across cycles \#9, \#10 and \#11.  
This broad line component is thought to be formed in the inner regions 
of the system, since it requires a
relatively high flux of energetic photons and high density. We do not know the exact location, 
but possible candidates are the walls of the WWC shock-cone and the inner wind of the primary star.
We must keep in mind that the central object is seen under heavy circumstellar extinction 
(A$_{V}\sim$7 mag) as compared to the Weigelt objects (A$_{V}\sim$0.5), which are primarily 
affected only by interstellar extinction \citep{hillieretal01}. Variability in these two components 
need not be correlated.

\begin{figure*}
\vbox {\vfil
\resizebox{17cm}{!}{\includegraphics{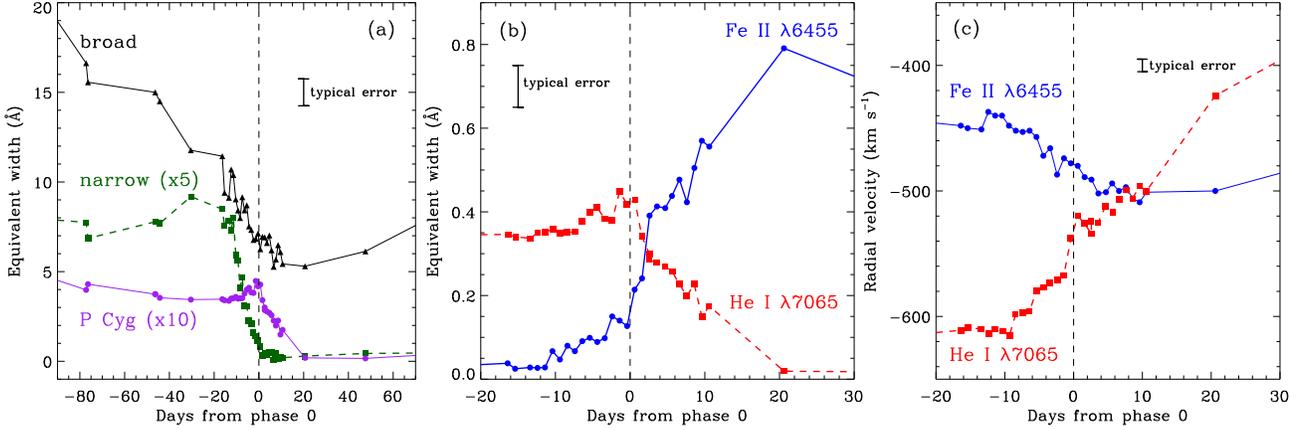}}
\caption{(a)  \label{he70} He\,{\sc i}~$\lambda7065$: equivalent width of the broad 
emission ({\it triangles}); narrow emission component multiplied by 5 ({\it squares}); and the 
P Cygni absorption component multiplied by 10 ({\it circles}); (b) Equivalent width of He\,{\sc i}~$\lambda7065$
({\it squares})and Fe\,{\sc ii}~$\lambda6455$  P Cygni absorption components ({\it circles}); 
(c) Radial velocity of barycenter of He\,{\sc i}~$\lambda7065$ ({\it squares}) and Fe\,{\sc ii}~$\lambda6455$ 
 P Cygni absorption components ({\it circles}). Typical error bars are shown in the upper 
part of the panels.
}
\vfil}
\end{figure*}

The central part of the minimum in this broad He\,{\sc i} line component 
shares some similarities with the narrow component, but there are also important differences.
Large variability is confined to a relatively short time interval  ($\phi$ = 0.99-1.035 or $\sim$90 days), 
very similar to the X-ray minimum. The centre of the minimum 
($\phi$ = 0.015 or $\sim$30 days after $\phi$ = 0) is also coincident with that in X-rays 
and occurs much earlier than in the {\it slow variation} component ($\sim$ 140 days). 
The minimum in the He\,{\sc i}~$\lambda6678$ broad
 component is a little asymmetric, in the sense that the fading is faster than
the recovering, but to a much smaller extent than for the narrow line formed in the Weigelt objects.

It is interesting to note that the broad emission never disappears, which could indicate that there is always some ionizing radiation from the secondary
star illuminating the walls of the shock-cone. Alternatively, the residual broad
emission could simply be `intrinsic' emission from the primary wind. 
The fact that the broad He\,{\sc i} component is formed near the center of mass and never disappears, 
indicates that eclipses cannot be invoked to explain all phenomena. For this line the real cause of the 
fast drop is a crash in the structure of the  wind-wind collision which may also affect 
the escape of high energy photons. Observations by \citet{b19} show that the broad line components
 weaken simultaneously in the Homunculus  south pole (via reflected light).
This rules out an eclipse (or `cone eclipse') mechanism, since different directions are affected simultaneously.  
We thus favor a model in which there is a collapse and restoration of the wind shock-cone. 
A beautiful realization of  this idea is shown in the 3-D numerical simulation by \citep{Okasaki}.

What is causing the changes in the WWC? The secondary star does not seem to be the culprit.
The repeatability of the X-ray light curve \citep{b2} is remarkable. Since the X-ray emission is dominated
by its wind (with the primary wind playing the role of a wall) the companion
must be a stable star. The change should be in the primary star. If its mass loss 
decreases, the WWC shock is shifted farther from the secondary star and the shock-cone aperture 
is enlarged. This would cause a decrease in the ionization and as a consequence a weakening of 
the  He\,{\sc i} broad line component. 

The WWC should reorganize some time after periastron, and its signature 
could be  present in our data. The fast recovery in the X-ray light curve and the He\,{\sc i} 
broad component line intensity at $\sim$3 months after $\phi=0$ indicates that at that time,
the WWC is already restored and that our  line-of-sight is again inside the shock-cone opening. 
Three months thus provides an upper limit to the duration of the WWC crash since
we cannot exclude the possibility that the recovery in 
the  X-ray light curve and the He\,{\sc i} broad component corresponds to 
the end of  the `cone eclipse'  to our specific direction. 


The `collapse' component takes much longer to end for the Weigelt objects. 
The ionization/excitation in the Weigelt objects (narrow components 
in the high ionization forbidden lines and in He\,{\sc i}) takes $>$9 months to reach the same 
level as it had a month before  $\phi=0$. 
This is possible since they are at a line-of-sight different from ours. More importantly, 
they are far from the central source and the external parts of the shock-cone are 
distorted as it rotates around the orbit.  After the collapse, additional gas may be trapped 
for some time inside the shock-cone. The slow recovery seen in the radio light curve
may be due to the same cause. 

\subsection{The {\it collapse} component in the P Cygni absorption lines}\label{PCygni}

P Cygni absorption lines are important because they sample the gas in a narrow beam 
to our side of the emitting region. They change as the source moves in the binary orbit and/or because
the absorbing material suffers  changes in the degree of excitation. Many lines display these
components, which are strongly 
variable across the minimum. However, changes from cycle to cycle have also been observed.

The P Cygni absorption component,  that is lost in He\,{\sc i}~$\lambda6678$ due
to a blend with [Ni\,{\sc ii}], appears clearly  in He\,{\sc i}~$\lambda7065$ 
line (Fig. \ref{he70}a)\footnote{The narrow and the broad emission
components of He\,{\sc i}~$\lambda7065$  display a behavior very similar to He\,{\sc i}~$\lambda6678$.}. 
It reaches EW$\approx$0.5~\AA\ at $\sim$3 months before $\phi=0$ 
and decreases with time. Three weeks before $\phi=0$  it
starts increasing, reaching a new maximum around $\phi=0$ and then it decreases up to complete 
disappearance $\sim$20 days after. While the P~Cygni absorption must be produced by material
between us and the primary star there is still a debate whether it is directly related to material
in the WWC cone, or produced by the wind of the primary \citep{krister07}. For either scenario,
the absence of P~Cygni absorption during the minimum indicates
that the aperture of the WWC cone is pointing away from us, which coincides with the position of
the secondary star `behind' the primary. With this orientation, He along 
our line of sight to the primary cannot be ionized by energetic photons from the secondary star. 
This gives support to the idea that the {\it collapse} component is produced
by an `eclipse-like' phenomenon that it is centered at $\sim$3 weeks after $\phi=0$.

The idea that the shock-cone is pointing away from us during the minimum 
is reinforced by comparing  the P Cygni absorption component of He\,{\sc i}~$\lambda7065$  with 
that of Fe\,{\sc ii}~$\lambda6455$. The large difference 
in excitation energy makes these two lines strategic to probe material 
at large and short distances from the ionizing source.
 In Fig. \ref{he70}b we see 
that across the minimum, the absorption component in Fe\,{\sc ii} is rising while 
that of He\,{\sc i} is fading, a behavior also observed by \citet{krister07} in HST 
spectra of the central source.  The observed behavior is a consequence 
of Fe\,{\sc ii} recombining on our side of the  primary star wind, that is located on the opposite side of the
primary star relative to the ionizing 
companion star. Just after the collapse, the wind of the primary star has recombined at 
maximum extent. It is interesting to see that the re-ionization 
(intensity decrease) of Fe\,{\sc ii} starts 
soon, as compared with the long absence of P Cygni absorption in He\,{\sc i}. 
This behavior of the He\,{\sc i} P Cygni absorption component is expected, since the opening of
 the shock-cone takes time to point again to our direction. The re-ionization 
of  Fe\,{\sc ii} in our side of the primary star wind, however, starts soon after the opposition 
of the secondary star, as its ionizing flux penetrates through the wind. The
 broad line component of  He\,{\sc i}~$\lambda6678$  in Fig. \ref{he66}c shows 
that the ionization recovers relatively quickly in the inner parts of the system, as compared to  
the slow recovery seen by the Weigelt objects.

The radial velocity of the P Cygni profile from these two lines provides an alternative way to analyze 
the situation just described. Fig. \ref{he70}c 
shows the same opposite behavior between He\,{\sc i}~$\lambda7065$  and 
 Fe\,{\sc ii}~$\lambda6455$ radial velocity as is shown by the line strengths. 
 The Fe\,{\sc ii}~$\lambda6455$ P Cygni absorption component shifts to more blueward velocities 
across $\phi=0$, while
the He\,{\sc i}~$\lambda7065$ P Cygni absorption component shifts to less blueward velocities.  
The amplitude of velocity changes is smaller in Fe\,{\sc ii}~$\lambda6455$ because
it is formed in regions where the wind is already reaching the terminal velocity
as compared to  He\,{\sc i} that is formed in a zone subject to larger acceleration. 
The maximum speed of the He\,{\sc i}~$\lambda7065$ P Cygni absorption component  
($-610$ km s$^{-1}$) is higher than that of  the Fe\,{\sc ii}~$\lambda6455$ P Cygni 
absorption component ($-500$\,km\,s$^{-1}$ ). Since Fe\,{\sc ii} recombination occurs 
in regions where the primary star wind is reaching its terminal velocity, the  
He\,{\sc i} line cannot be formed in the same stellar wind, as it requires higher 
excitation flux and gas density, only present close to the star.
The only plausible formation region for the  He\,{\sc i} broad line are the walls 
of the WWC shock-cone, not the primary's wind.

\begin{figure*}
\vbox {\vfil
\resizebox{17cm}{!}{\includegraphics{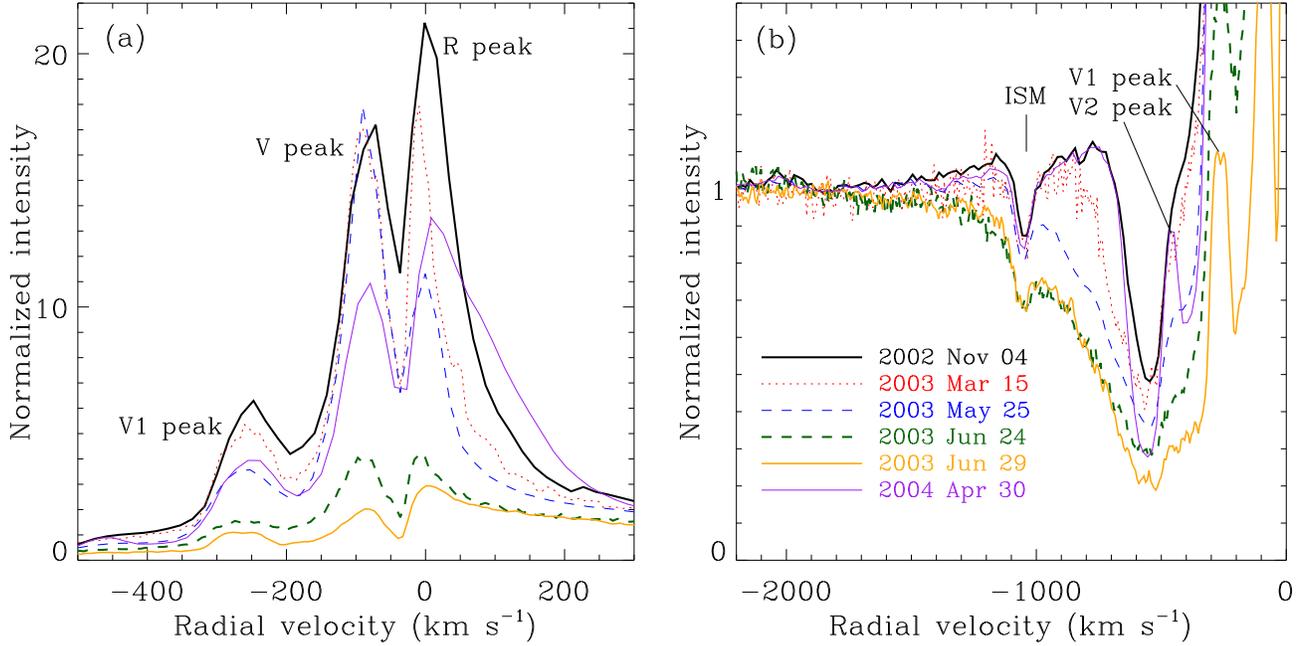}}
\caption{\label{10830} 
 Line profile of He\, {\sc i}~$\lambda10830$. (a) Variability of the emission peaks 
(legend in panel b).
(b) Zoom showing variations of the P Cygni absorption. Notice the extension of the 
profile up to $-1700$~km~s$^{-1}$ near $\phi=0$.}\vfil}
\end{figure*}

\section{He\,{\sc i}~$\lambda10830$~\AA - a peculiar spectral line}\label{hei10830}

The He\,{\sc i}~$\lambda10830$ line is particularly interesting: it is very strong,  displays 
many components in the central knot and in the nebular region, and 
undergoes remarkable variations.
Figure \ref{10830}a displays the variability of the emission profile. 
It changes in intensity and width along the cycle. At high resolution,
two peaks are clearly defined +12 (the `R' peak)  and $-85$\,km\,s$^{-1}$ (the `V' peak).
The `R' peak is higher  than the `V' peak (R/V$>$1) during the high excitation state.
The intensity ratio changes to R/V$<1$ at 105 days before $\phi=0$ and 
again to R/V$>$1 at 4 days before $\phi=0$ (Paper I).
There is 
a question if the V and R peaks are independent emissions, or just a broad emission split in two 
by a shell absorption feature, as claimed by \citet{b29}.  Our present high resolution spectra 
show that the putative `shell absorption' does  fall below the continuum 
only during the minimum. This is a false continuum,  due to the depression of 
the local stellar continuum by the P Cygni absorption. 
The pseudo absorption profile remains steady at $-35$\,km\,s$^{-1}$, 
as also the  V and R peaks do.  Although the radial velocity of the narrow absorption
feature agrees with the faint absorption in Balmer lines, the R and V peaks vary 
independently and are steady in radial velcoity indicating separate emitting regions. 
The Weigelt objects are natural candidates
for these two components, but  the large separation in radial velocity ($97$\,km\,s$^{-1}$) does 
not match that of the Weigelt objects. Moreover, when looking to long slit spectra, we see 
two narrow emission lines crossing all the field upt o the borders of the Homunculus. The 
two nebular lines are separated by $-127$\,km\,s$^{-1}$, a little more than the 
R and V peaks.  This happens because, in the central object, the nebular lines merge with the 
 R and V peaks that are much broader and contaminated by the velocity field 
of the circumstellar gas.

Figure \ref{10830}a shows  two additional emission peaks  at $-255$ (V1 peak)   
and $-460$\,km\,s$^{-1}$ (V2 peak). These two peaks are variable along the cycle and 
also from cycle to cycle. 
The  absorption component at $-$1060~km~s$^{-1}$  (ISM) was once believed
to be a narrow absorption component in the stellar wind \citep{b31}, 
but \citet{b90} showed that it is an interstellar/circumstellar feature.
The P Cygni absorption during the high excitation phase 
is faint and placed at  $-$570~km~s$^{-1}$, in agreement with the lines
emitted by the wind of the primary star. As the minimum
approaches, this feature starts to get broader and deeper, 
the centroid shifts up to $-$650~km~s$^{-1}$.  This change in the centroid is due to the 
contribution from the absorption wings that increases, extending up to $-$1400~km~s$^{-1}$.
When plotting together the spectra at maximum and minimum excitation, 
it is possible to track the line wings up to  $-$1700~km~s$^{-1}$
While high velocities are generally not seen at our viewing angle,
velocities of order 1000~km~s$^{-1}$ are
seen from polar directions \citep{b33}. This led \citet{b33} 
to suggest that the wind of $\eta$~Car is asymmetric, and that during the
event a shell event occurs leading to a situation in which the
equatorial wind more closely resembles the polar wind.

The P Cygni absorption component in He\,{\sc i} $\lambda10830$   
reaches its maximum strength (EqW = 18 \AA) 10 days after  $\phi=0$, a time when
the absorption components in the optical He\,{\sc i} lines have already faded. 
This could indicate that different spatial locations contribute to the observed absorption,
although all absorbing structures must occur in gas between us and the primary.
Surprisingly, the variability of the P Cygni line is more similar to the low excitation 
line  Fe\,{\sc ii}~$\lambda6455$ and that of He\,{\sc i}~$\lambda7065$.
An alternative explanation for the distinct behavior of the absorption component of He\,{\sc i} $\lambda10830$
is related to the stability of the 2s $^3$S state. After the ionizing
radiation field is switched off, the population of the metastable  2s $^3$S state will persist longer
than the 2p states, which are the lower levels for the observed optical He\,{\sc i} lines.

The radial velocity of the P Cygni absorption component in He\,{\sc i}~$\lambda10830$ 
has a behavior similar to  that of He\,{\sc i}~$\lambda7065$, with the minimum (maximum negative) velocity 
occurring  $\sim$ 10 days before  $\phi=0$ and a maximum at  $\sim$ 45 days after.
The maximum speeds in He\,{\sc i}~$\lambda10830$ and He\,{\sc i}~$\lambda7065$ are 
$-$650, $-$610~km~s$^{-1}$ and $-$450 and $-$350~km~s$^{-1}$  respectively.
The minimum (maximum negative) speed of the He\,{\sc i}~$\lambda10830$  P~Cygni line 
is also larger than that of Fe\,{\sc ii}~$\lambda6455$, indicating that it is not formed in the 
outer wind of the primary star. 
One possibility is that He\,{\sc i}~$\lambda10830$ is forms in gas from the secondary 
star that has passed through  the shock-cone, forming a  tail  left behind in a spiral like pattern,
as the secondary star turns around the center of mass.  When such portion of high velocity gas ceases to
be ionized by its parent star, it recombines, producing the  extended wings in the  P Cygni absorption.

\begin{figure*}
\vbox {\vfil
\resizebox{17 cm}{!}{\centering \includegraphics{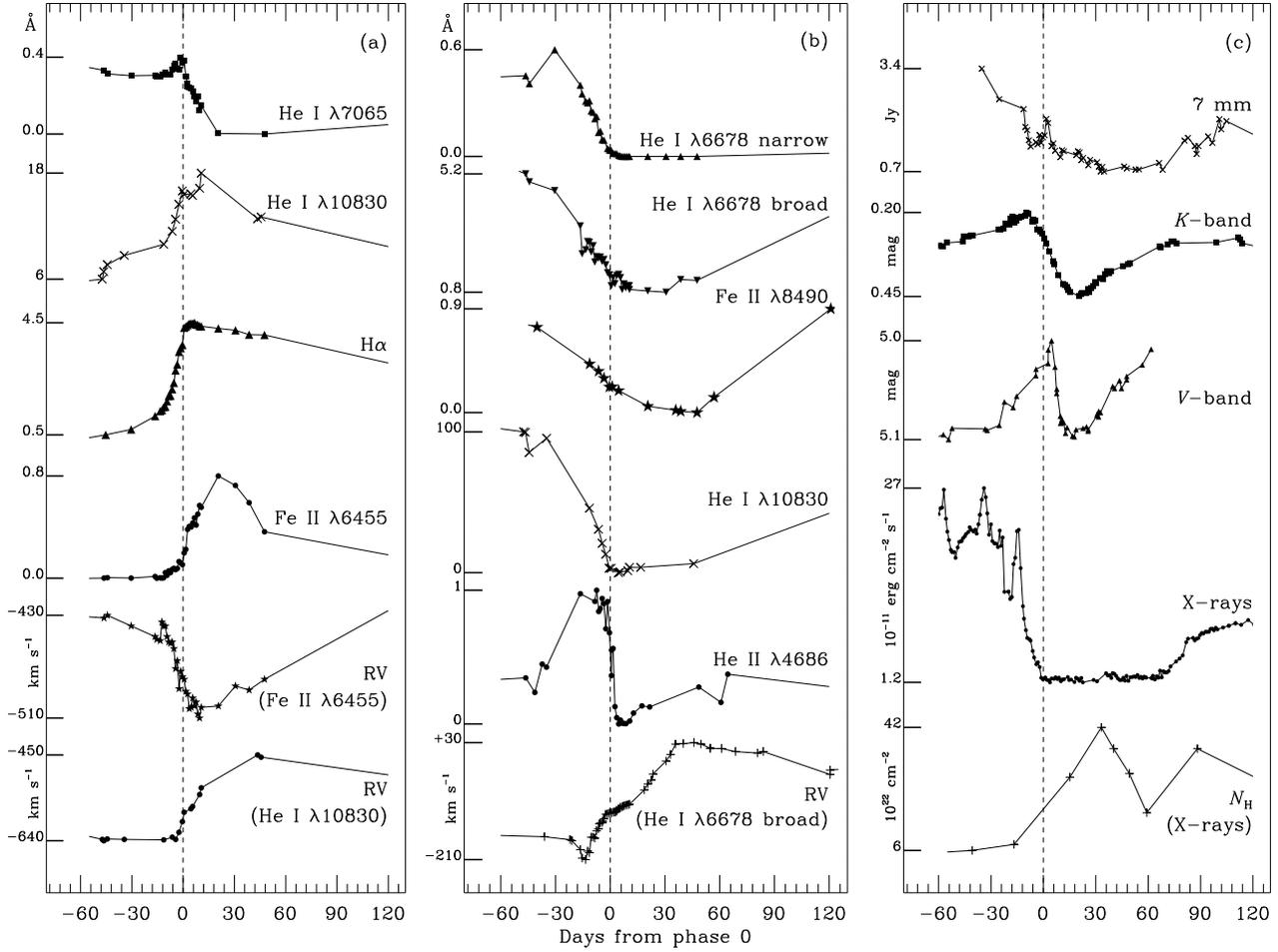}}
\caption{\label{2009} A panoramic view of variations in different features,
and at different wavelengths, in  the core of the event ({\it collapse} component). 
(a) P Cyni absorption. EqW in ~\AA: He\,{\sc i}~$\lambda7065$;
He\,{\sc i}~$\lambda10830$;  H$\alpha$ and  Fe\,{\sc ii}~$\lambda6455$ plus
radial velocity of Fe\,{\sc ii}~$\lambda6455$  and He\,{\sc i}~$\lambda10830$ in~km~s$^{-1}$. 
(b) Emission lines.  EqW in \AA:  He\,{\sc i}~$\lambda6678$ narrow and broad component;  
Fe\,{\sc ii}~$\lambda8490$ (narrow) fluorescent line; He\,{\sc i}~$\lambda10830$ total 
line profile;  
He\,{\sc ii}~$\lambda4686$ \citep{b20}; and radial velocity of He\,{\sc i}~$\lambda6678$ 
broad emission component in~km~s$^{-1}$. 
(c) Broad band. Radio 7-mm flux in Jansky \citep{b1}; $K$-band mag \citep{b24}; $V$-band 
mag \citep{b14}; X-ray flux in 10$^{22}$  erg cm$^{-2}$ s$^{-1}$ \citep{b2};  and X-ray column 
density ($N_H$)  in 10$^{22}$ cm$^{-2}$ \citep{b11}.}
\vfil}
\end{figure*}

The observed behavior of the He\,{\sc i} $\lambda10830$ emission profile
is a combination of several effects. First, around the event, 
the Weigelt objects are shielded by the dense primary wind
(and perhaps the star itself) from the ionizing
radiation field, causing the narrow component to weaken, and disappear
entirely during the state of minimum. Second, in order
to explain the weakening of the broad component, the size of the
He\,{\sc i} emitting region must shrink near periastron.
Since the number of ionizing photons emitted by the secondary star
does not change during the event, why does the broad He\,{\sc i} emission
change? This must be due
to the dense primary wind reprocessing the ionizing photons, and perhaps
even the emitted He\,{\sc i} photons. This in turn requires that the secondary star
be deep within the primary wind at periastron. 
The residual broad He\,{\sc i} emission, in both $\lambda10830$ and optical lines,
that is observed during the minimum  is probably
produced by the primary. One possible caveat with this explanation is the weakness of the
P~Cygni absorption after the minimum.
The calculations of \citet{hillieretal01} show that it is relatively easy to
alter the strength of the intrinsic He\,{\sc i} emission without significantly
affecting H and Fe\,{\sc ii} line strengths, and thus this residual 
He\,{\sc i} emission is easily explainable for reasonable LBV parameters.
One final feature that requires explanation is the high velocity absorption
profile seen during the minimum. 

At minimum, the observed profile closely resembles that of HD 151804, which is an
O8Iape \citep{b90}. This must occur simply by chance, and cannot be
the secondary star, since \citet{HGN06_UV} showed that we would not 
expect to see the companion spectrum in the optical/IR,
since the primary's luminosity is at least a few times higher than the secondary.

\section{A multispectral view of the minimum}\label{multispectral}

The spectroscopic events are present at all wavelengths --
radio, IR, optical, UV, and X-ray -- and in spectral features
such as the doubly ionized forbidden lines, the narrow He\,{\sc i} lines, 
the broad He\,{\sc i} lines, and the broad emission lines from the wind of the primary star. 
While all show the
event, the characteristic behavior varies dramatically. At one extreme, 
the radio 3-cm light curve  shows only the {\it slow variation} and at the other
extreme, He\,{\sc ii} and X-rays show only the {\it collapse} component.
The observed behavior rules out any mechanism for the event  that works only close to
periastron, such as a shell ejection around periastron \citep{b15}, accretion of the 
primary wind onto the secondary companion  \citep{soker05} and 
eclipse of the shock-cone opening \citep{b16}. Models like these, in which 
the minimum is produced only by geometrical factors (`cone eclipses') are 
ruled out by radio observations.

In Fig. \ref{2009} we present a panoramic view  of emission, absorption, radial 
velocity, and continuum flux variations across the minimum (collapse component), 
and across the spectrum. In addition to the measurements made in our spectra, 
we reproduce curves by other authors, taken from the cited literature.
Here, we are not aiming to present full data, just the light curves morphology, 
to show the time of their maxima/minima and the decline/recovery phases.
Since the plots are compressed in the Y scale, the errors are 
of the order of the size of the symbols.

We want to warn the reader that radial velocities presented were 
derived from the position of the centroid of broad lines. The line profiles are complex, 
and thus radial velocity variations do not represent the movement of  some specific region. 
Our aim, by measuring centroids, was to produce robust model independent measurements. 
Variable and extended  line wings introduce shifts in the centroids, 
not only in proportion to the wing extension, but to its intensity. In this work, we are not aiming to measure the full 
extent of the line wings, which is difficult because of their assymptotic merging with the 
stellar continuum. Such measurements would rely on subjective definition of  the terminal speed, in
addition to being limited by the signal-to-noise of the stellar continuum.  Centroids represent the bulk 
of the velocity field.

The top 4 curves of Fig.  \ref{2009}a display the
EqW variation of P Cygni absorption features for four lines -- each line shows its own unique
variability curve. 
The EqW of He\,{\sc i}~$\lambda7065$ P Cygni absorption is generally high along the 5.5-yr cycle, 
reaching a peak 2 days
 before $\phi=0$. It then decreases with time, reaching the minimum 20 days after $\phi=0$. 
 In contrast, the EqWs of the P Cygni absorption components of He\,{\sc i}~$\lambda10830$,  H$\alpha$ and
Fe\,{\sc ii}~$\lambda6455$ are low before $\phi=0$, reach the maximum after that, 
and maintain a high intensity level for a relatively long time after $\phi=0$.
  H$\alpha$ reaches its maximum during the first week after $\phi=0$ and seems to be saturated,
decreasing very slowly after that. He\,{\sc i}~$\lambda10830$ reaches 
its maximum 10 days after $\phi=0$, while Fe\,{\sc ii}~$\lambda6455$ takes 20 days 
to reach its maximum.

The radial velocity variations of the  Fe\,{\sc ii}~$\lambda6455$ and 
He\,{\sc i}~$\lambda10830$ P~Cygni absorption components  are presented at the 
bottom of Fig.~\ref{2009}a. We measured the position of the line centroid, by integrating along 
the absorption line profile, and so it corresponds to the barycenter of the P Cygni absorption component. 
Starting $\sim$2 months before $\phi=0$, Fe\,{\sc ii}~$\lambda6455$ 
radial velocity shifts to higher (negative) values, reaching a minimum $\sim$10 days after $\phi=0$ 
 and returns to pre-minimum at a slower pace.
The centroid of the P Cygni absorption component shifts by $\sim$80~km~s$^{-1}$ across 
$\phi=0$, indicating that the bulk of the recombination shifts toward larger radii.
We have already shown that He\,{\sc i}~$\lambda7065$ P Cygni absorption
moves in the opposite 
sense across $\phi=0$, changing by 250~km~s$^{-1}$ in the time frame 
starting 10 days before $\phi=0$  up to  50 days after it. The radial 
velocity of the He\,{\sc i}~$\lambda10830$  P Cygni absorption component 
behaves like He\,{\sc i}~$\lambda7065$.
This seems confusing, since the intensity of this line behaves like  Fe\,{\sc ii}~$\lambda6455$.  
This can be explained if the He\,{\sc i}~$\lambda10830$ line  
has contributions from different regions, in the inner and outer 
parts of the system. The core of the P Cygni absorption in  He\,{\sc i}~$\lambda10830$ 
seems to be formed close to that of He\,{\sc i}~$\lambda7065$, in the inner regions 
of the shock-cone and primary star wind. The wings of the P Cygni 
absorption, however, can be formed at larger distances, contributing substantially 
to the EqW, but not too much to the radial velocity. We will discuss this point later 
in this paper.

We now turn our attention to emission line
strengths, displayed in Fig. \ref{2009}b.
The He\,{\sc i}~$\lambda6678$ narrow component emission behaves 
like the doubly ionized forbidden lines that are emitted in the Weigelt objects; 
there is an extended minimum after  $\phi=0$  and a very slow recovery. The broad
component of the same line shows a minimum at a later time 
and a faster recovery than the narrow component. 
A narrow fluorescent line, Fe\,{\sc ii}~$\lambda8490$, shows a
minimum at an even later time, but the general appearance of the {\it collapse} component
is very similar to that of  the He\,{\sc i}~$\lambda6678$ broad component. 
The EqW of the complex He\,{\sc i}~$\lambda10830$  line displays a behaviour intermediate 
between that of the narrow and the broad components of 
He\,{\sc i}~$\lambda6678$, indicating that its emission line  profile is composed of both
narrow and broad components.

The He\,{\sc ii}~$\lambda4686$ line intensity curve is taken from \citet{b20}. 
This line remains absent from the spectrum most of
the cycle \citep{b20,b15};  rises quickly the month before  $\phi=0$ and displays a very sharp 
decline after that. 
The same behavior was reported by \citet{b19} in the central star and reflected by dust 
in the pole of the Homunculus SE lobe: 
at 2.6~arcsec south and 2.8~arcsec east  from the star.
A narrowing of the shock-cone, as in the case of X-rays, could
explain the disappearance of He\,{\sc ii} from our line-of-sight and
simultaneously (after correcting for the light time travel) from the Homunculus SE pole, 
as long as the cone opening is pointing away from us at periastron, in agreement
with other observations. \citet{b20} argued that He\,{\sc ii} $\lambda$4686
was excited by X-rays from the WWC, but \citet{b15}
argued that this mechanism was too inefficient, and instead invoked
a mechanism related to UV radiation from the central source although
the precise details are unclear. The simultaneous disappearance 
from multiple viewing angles plus the strongly asymmetrical light curve
suggest that He\,{\sc ii} 4686 emission is not simply a result
of UV excitation in the inner wind of the secondary star, but instead might 
somehow be coupled to the WWC whose shape is strongly affected around periastron.

At the bottom of Fig.  \ref{2009}b we present the radial velocity curve of 
the He\, {\sc i}~$\lambda6678$ broad emission component. It shows a minimum 
of $\sim-210$~km~s$^{-1}$ at $\sim$2 weeks before $\phi=0$, a maximum (30~km~s$^{-1}$)
in coincidence with the middle of the X-ray minimum and then a slow decrease along 
rest of the 5.5-yr cycle.

Fig. \ref{2009}c shows measurements in the broad continuum bands. 
The light curve at $7$-mm radio wavelength was made with data from \citet{b1}.
It is interesting, since the shape of the minimum is between that of long radio 
wavelength ($3$-cm), which varies slowly along the whole cycle and the
of X-ray light curve, which has a very sharp minimum.
The middle of the radio-mm minimum occurs $\sim$2 weeks after that in X-rays. 

The light curve in the K-band was taken from \citet{b24} and the V-band from \citet{b14}.
The center of the minimum in the V and K bands occur $\sim$2 weeks before that in X-rays. 
Further, the duration of the minimum in these two
photometric bands is the shortest among all the measured features. Both of them show 
a peak preceding the minimum, as discussed by \citet{b24}, and in both cases the variability
amplitude is small ($\Delta$V= 0.1,  $\Delta$K=0.25).

The variation of $N_{\rm H}$, in units of 10$^{22}$~cm$^{-2}$ and reported by \citet{b2}, is presented at 
the bottom of panel \ref{2009}c. It behaves similarly to  the Fe\,{\sc ii}~$\lambda6455$  P Cygni 
absorption. It is important to notice that the time of maximum in these two features coincides
with the middle of the X-ray minimum. We suggest that this time
corresponds to the periastron passage.
There is a  secondary peak at $\sim$3 months after $\phi=0$, but 
we do not know if it is real, or just an unreliable measurement. 
 \citet{b11} show that the temperature
of the WWC remains constant through the minimum, in contrast
to the behavior expected if the shock cools and collapses.
But \citet{b2} \& \citet{b11} show that there is a reduction
in the emission measure during the minimum, which means
that almost all of the WWC is hidden behind a very thick
absorber. There is also some indication from the X-ray line
profiles for a significant change in the physical condition in
the WWC. One possibility would be that the shock-cone
narrows significantly, in order that X-rays escape only to
some defined direction. In this way, the X-ray light curve
is modulated by both intrinsic and extrinsic (geometrical)
causes.

\begin{figure}
\vbox {\vfil
\resizebox{8.4 cm}{!}{\centering \includegraphics[width=0.45\textwidth]{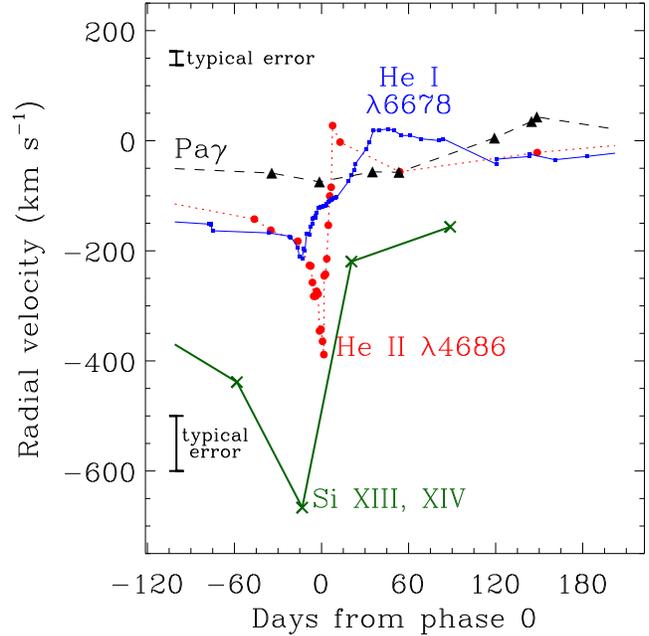}}
\caption{\label{RV} 
Radial velocity curves of broad emission lines close to $\phi=0$, 
covering a wide range of  excitation energies. Si\,{\sc xiii-xiv} lines were 
averaged from data reported by \citet{henley08} and He\,{\sc ii}$\lambda4686$ 
was taken from \citet{b20}. Tipical error bar for the optical/IR lines 
are in the upper part of the figure and for the combined X-ray lines at the bottom left.
}\vfil}
\end{figure}

In Figure \ref{RV} we present radial velocity curves for broad lines covering 
a wide range of excitation energy: the broad emission lines of Si\,{\sc xiii-xiv} 
- taken form \citet{henley08},   He\,{\sc ii}~$\lambda4686$ - taken from \citet{b20},  
He\,{\sc i}~$\lambda6678$ and   H\,{\sc i}~$\lambda10938$ and the 
 He\, {\sc i}$\lambda7065$ absorption, measured by us. Si\, {\sc ii}~$\lambda6347$ broad emission and P Cygni 
absorption  behave like He\, {\sc i} (not shown here).  We see a common signature of a slow shift toward  
higher negative speeds when approaching $\phi=0$, then 
a sudden reversal to the positive side, followed by a new slow decrease toward the negative side.
The extreme observed radial velocities are 
$-$700~km~s$^{-1}$ for the X-ray Si\,{\sc xiii-xiv} lines \citep{henley08},  $\sim$ $-$400~km~s$^{-1}$
for He\,{\sc ii}, $\sim$ $-$200~km~s$^{-1}$ for He\,{\sc i} and $\sim$ $-$80~km~s$^{-1}$ for
H\,{\sc i}. The velocity jump at  $\phi=0$ is higher the higher the excitation energy. 
Although the radial velocity changes are due to binarity, they do not track the orbital motion  of the 
companion stars. \citet{henley08}  were able to fit the X-ray lines if they are formed in 
the WWC cone walls. 

Could the lower excitation lines presented in Fig \ref{RV} be interpreted 
in the same way as X-ray lines, since they share similar properties?
 He\,{\sc ii}~$\lambda4686$, for example, is believed to be formed close to the X-ray emitting region 
\citep{b20,b15}. However,  it is formed at lower temperatures than X-ray lines, which means 
farther from the apex than X-ray lines. Since the gas speed increases as it flows away from the 
cone apex, the lower excitation lines should have higher speeds, contrary to that 
is  observed. The fact that the P~Cygni absorption in He\,{\sc i}~$\lambda7065$ 
follows the same general pattern as very high energy emission lines is even more 
difficult to understand. One possibility is that, far from the apex,  the conical surface does not 
exist anymore or it is dominated by large disturbed cells.  

\section{ DISCUSSION AND CONCLUSIONS}
\label{discus}

From an analysis of the narrow components of forbidden  emission lines we infer that
the event is composed of two parts: the {\it slow variation}, and the 
{\it collapse} components.  The former comprises almost the entire 5.5-yr cycle, 
is centered at a later time than the {\it collapse} component and the fading/recovering 
branches have a smaller degree of asymmetry. 
These lines are formed in the Weigelt objects and 
represent the way they `see'  the ionizing source, which is the radiation 
field of the secondary star that escapes through the shock-cone opening. 
The gentle fading and recovery indicates that the cone is not completely 
transparent. The obscuration of the ionizing source, as seen from the
Weigelt objects, increases and decreases as the secondary star 
travels around the orbit and deep into and out of the primary's wind. The 
{\it slow variation} component indicates that the shape and gas density 
inside the shock-cone change along the cycle. 
Models invoking a shell ejection, accretion onto the secondary  companion, 
or a simple eclipse by the primary, are unable to explain the {\it slow variation} component.

The {\it collapse} component is seen in many spectral features ---
X-rays, high excitation Weigelt lines, He{\,\sc i} P~Cgyni absorption, and 
He{\, \sc I} emission. It is likely that multiple processes are affecting the
collapse phase, with its duration ultimately set by the amount of time the
companion spends near periastron in its highly eccentric orbit. 
Eclipse-like effects (`cone eclipse') could produce a decrease in lines formed in the Weigelt 
objects, in the inner system and reflected in the dusty Homunculus and features. 
However, not all directions can be affected at the same time. 

Due to uncertainties in the orbital and stellar parameters
it is unclear how close the primary and secondary stars approach at periastron.
However, because of the much higher densities near periastron, and
the rapidly changing orbit, conditions in the shock-cone will change, and it is
 possible that instabilities will occur.
The observed behavior of many spectral features might be explained  
by a global disruption of the wind-wind collision shock.
With this model it is also easy to explain the observed 
asymmetry seen in the {\it collapse} light curves around minimum.
The bow shock is not symmetrically orientated
around periastron and its structure is not identical either side
of periastron. In addition to the Coriolis effect, the 
companion stars are approaching each other before periastron 
and receding after, changing the relative speeds at which the winds collide.

Although we could not identify unambiguously the  signature of the WWC restoration, 
it must happen at most 3 months later than $\phi=0$. At ~3 months, their is the
sharp recovery in the intensity of X-rays and the He\,{\sc i} broad emission lines.
This is either due to the restoration of the WWC, or due to the end of the 
`cone eclipse' (in which case the restoration of the shock must have occurred earlier).

The doubly ionized forbidden lines show a remarkable 
behaviour in the {\it collapse} component. 
There is a trend of the time delay of line fading 
with the  ionization potential, indicating that the Weigelt objects
see an ionizing spectrum that becomes progressively 
softer as the minimum is approached, and progressively harder 
in the recovery phase. This must be caused by optical depth effects
of intervening gas in between the ionizing source and the
blobs. The slow re-ionization of the blobs ($>$6 months 
after the minimum for
[Ne\,{\sc iii}]) is most likely due to obscuration effects.
Since the line intensity curves of the high energy forbidden  lines do not 
flatten out in the high excitation state, it is probable that 
only a fraction of the gas in the Weigelt objects is 
ionized to Ne$^{++}$,  S$^{++}$ and Ar$^{++}$.

Very high energy features,such as X-rays and  He\,{\sc ii}~$\lambda4686$, 
display almost only the {\it collapse} component. Conversely, at radio cm wavelengths 
the light curve is dominated by the {\it slow variation} component.
At mm wavelengths the {\it collapse} component is present at a modest level. 

The minimum of the radio 7-mm light curve is centered 
at $\sim$42 days after $\phi=0$. Although the WWC shock may be 
responsible for a fraction of the radio emission at these energies, 
most of the emission comes from ionization in the gas outside the 
binary system. Some effects, such as the light travel time from the source 
and recombination time of the gas,  can represent a delay to the 
center of the minimum in order that this epoch is compatible with the 
center of minimum in X-rays. 

He\,{\sc i}  P Cygni absorption (and broad emission) lines cannot be formed 
in the wind of the primary star, since its speed reaches values larger than 
those of  P Cygni absorption in Fe\,{\sc ii}~$\lambda6455$, which is 
formed in the external layers of the primary's wind.

As the P Cygni absorption component of Fe\,{\sc ii}~$\lambda6455$ 
reaches the maximum intensity and maximum negative velocity 
 $\sim$3 weeks after $\phi=0$. the wind of the primary 
star is recombined to a maximum extent at that time. It must occur when the 
 (ionizing) secondary star is at opposition. This interpretation is corroborated 
by the fact that around $\phi=0$ the velocity of the P Cygni absorption component of 
Fe\,{\sc ii}~$\lambda6455$  is shifting to the negative side, and 
that of He\,{\sc i}~$\lambda7065$ P Cygni absorption to the opposite direction. 
 Other features associated with the shock-cone of the secondary star 
support this scenario, such as the He\,{\sc i}~$\lambda6678$  broad
 emission component that also is shifting to the 
positive side of radial velocities and reaches the minimum at $\phi$ = 0.01. 
The X-ray flux reaches a minimum and the column density 
($N_{\rm H}$) reaches a peak 30 days after $\phi=0$. Moreover, 
the fluorescent (narrow) line Fe\,{\sc ii}~$\lambda8490$  also 
reaches a minimum around  $\phi$ = 0.015.  These data, taken together,
indicate that the opposition of  the secondary star occurs 
20-30 days after $\phi=0$. We do not claim this is an eclipse in a classical sense, 
neither from the primary stellar disk nor from its wind.

The He\,{\sc i}~$\lambda$10830 line indicates that material
along our line-of-sight latitude reaches much higher
negative velocities ($-$1400~km~s$^{-1}$ ) than the typical 
velocities we see in the
high excitation phases ($-$550~km~s$^{-1}$). 
Most likely the P Cygni absorption
is due to secondary gas that flows through the walls of the shock-cone
 and is left behind as the secondary star turns around the center of mass 
at periastron. Variations in the V and R line peaks indicate that 
there are other emitting regions for this spectral line, in addition to the 
Weigelt objects and the winds of the companion stars. 

Although the observations reported here are compatible with the generic 
binary scenario presented in section \ref{binary}, none of the existing models 
describes the whole data set. Numerical 3-D simulation of two-colliding
winds for parameters appropriate to the $\eta$ Carinae system will be needed to
assist in understanding the complex variability that is observed. 
Work in progress \citep{Okasaki} is very promising, having been able 
to reproduce the near-periastron collapse, compatible with 
data we presented here.

If we are to understand $\eta$ Carinae and its evolutionary state, the identification 
of genuine Keplerian velocity-shifts, and the determination of the 
masses and spectral types of both stars, is a crucial goal.
The intensive efforts applied to understand $\eta$~Carinae, which is a difficult system 
because of extreme conditions (masses, mass-loss, orbital eccentricity), brings 
benefits to the field of colliding wind binaries. Binarity in eta Car may not
have been responsible for the 1843 giant eruption, and so does not explain 
the other examples of extreme LBVs, like P Cygni, V1 in N2363, etc.
 However, the secondary star in $\eta$~Carinae probes different  regions of the primary's 
wind and, hopefully, will enable the determination of its mass.  This is 
helpful to understanding the parameters of evolved massive stars, which 
are difficult to diagnose when in isolation. 

The next minimum is predicted to start on 2009 January 11 ($\pm$2 d).  
Fig. \ref{2009} is useful to plan monitoring campaigns 
through the {\it collapse} component.
The time frame starts in early 2008 November and extends
through 2009 July. This is the best event since 1948 for
ground-based observations, since its central core fits entirely in the good
observing season. The next favorable event will not occur before
2020. The critical time for narrow components of the high excitation lines and
 He\,{\sc i} encompasses the time interval
December 15 to January 20. The P Cygni absorption and
the $N_{\rm H}$ to the X-ray source will reach the maximum around
February 10. In order to improve significantly the observations
made in the 2003.49 event it is necessary to monitor this
critical time interval more frequently than one observation
per day.

Since we now have a better picture of the events, we can plan
for the next critical observations focused on a few representative lines. 
Major gains, relative to the present knowledge, will come from higher 
spatial resolution and signal-to-noise observations. Signatures from high energy 
phenomena, like X-rays, He\,{\sc ii}~$\lambda$4686 and 
the Balmer jump  are especially important.

\section{Acknowledgments}
 We thank to J.E. Steiner, T. Gull and N. Soker for their comments on the draft. A.D., J.H.G. and M.T. 
thank FAPESP and CNPq for continuing support.

\bsp

\label{lastpage}

\end{document}